\documentclass{article}

\usepackage[nonatbib]{neurips_2018}

\usepackage[utf8]{inputenc} 
\usepackage[T1]{fontenc}    
\usepackage{hyperref}       
\usepackage{url}            
\usepackage{booktabs}       
\usepackage{amsfonts}       
\usepackage{nicefrac}       
\usepackage{microtype}      

\title{EEG-based Communication with a Predictive Text Algorithm}
\usepackage{graphicx}
\usepackage{subcaption}

\begin{document}

\maketitle
\begin{abstract}
Several changes occur in the brain in response to voluntary and involuntary activities performed by a person. The ability to retrieve data from the brain within a time space provides a basis for in-depth analyses that offer insight on what changes occur in the brain during its decision-making processes. In this work, we present the technical description and software implementation of an electroencephalographic (EEG) based communication system. We read EEG data in real-time with which we compute the likelihood that a voluntary eye blink has been made by a person and use the decision to trigger buttons on a user interface in order to produce text. Relevant texts are suggested using a modification of the T9 algorithm. Our results indicate that EEG-based technology can be effectively applied in facilitating speech for people with severe speech and muscular disabilities, providing a foundation for future work in the area.
\end{abstract}

\section{Introduction}
Communication is a vital aspect of life––a means of self-expression. Its importance is reflected in the vast availability of digital communication tools ranging from mobile phones and messaging applications to social networks and conversational assistants. As useful as these tools are, accessibility remains an issue of concern for people with disabilities. Mariusz'  \cite{duplaga}  study of internet usage in Poland found that people with disabilities experienced a significant digital divide - an inability to access the internet -  when compared to people without disabilities. Many existing communication software require typing on the keyboard (for text messaging) or speaking to a microphone (for voice messaging) for transmission to a second party. The need for alternative means of communication that adequately includes those who suffer from severe speech and muscular disabilities is therefore paramount. Roy et al. \cite{roy} show from their analysis of electroencephalographic (EEG) data that one can leverage the connections between the eyes and the brain to use eye blinks as a means of communication. In this work, we create a prototype of a reliable and easy to use EEG-based communication system for people who suffer from severe speech and muscular disabilities employing eye blinks as input modality. This approach offers an advantage over eye tracking cameras as cameras require constant focus on the eye to correctly track gestures.

\section{Related Work}
\label{gen_inst}

Over the past decades, a wide range of Brain-Computer Interface (BCI) applications have emerged. One of the classic goals of brain-computer interfaces are to unveil and utilize principles of operation and plastic properties of the distributed and dynamic circuits of the brain. Creating new therapies to restore mobility and sensations to severely disabled patients is another vital goal of BCI \cite{lebedev}.

Nagels-Coune et al. \cite{nagels} implemented a yes/no communication paradigm that relies on mental imagery (mental drawing) and portable functional near-infrared spectroscopy. In their study, participants either performed mental drawing, to encode a 'yes', or did not change their mental state, to encode a 'no'. Decoding accuracy reached 70\% in almost half of the participants. 

In a survey administered to people with motor-neuron disease in New South Wales, 15\%-21\% had difficulty with speaking, writing and/or using a keyboard. Majority of the subjects indicated a preference for portable solutions like laptops, tablets, mobile phones. A third of respondents used Skype or its equivalent, but few used this to interact with health professionals \cite{mackenzie}.

In a usability study carried out by Holz et al. \cite{holz}, a BCI prototype was evaluated in terms of effectiveness (accuracy), efficiency (Information Transfer Rate (ITR) and subjective workload) and users’ satisfaction by four severely motor restricted end-users; two were in the locked-in state and had unreliable eye-movement. Although all 4 subjects accepted the BCI application, rated satisfaction medium to high and indicated ease of use as being one of the most important aspect of BCI applications; sources for dissatisfaction were low effectiveness, time-consuming adjustment and difficulty of use.

\section{Method}
In the following subsections, we present the various components in our implementation of the BCI system.

\label{headings}
\subsection{EEG Data Capturing}
The NeuroSky MindWave headset used in Katona et al. \cite{katona} is used to capture the EEG signals. The headset has two dry EEG sensors; one is placed on the forehead while the other is clipped to the right ear lobe. The headset filters out extraneous noise and electrical interference, then converts the signal into digital power which is continuously transmitted via Bluetooth connection to the BCI application \cite{joyce}.

\subsection{Data Transfer}
In order to establish communication between the EEG application and the headset, a communication driver, which takes a communication (COM) port number and baud rate as inputs, is used to establish a communication link. Upon the selection of an appropriate communication port and baud rate, the application is able to continuously capture EEG signals \cite{eeg} which include the estimated eye blink strength and the delta, alpha, beta, and gamma values classified according to their frequency bands.

\subsection{Data Processing} 
Eye blink strength parameter is continuously monitored to know when an eye blink is made. An eye blink is said to be voluntary when the eye blink strength goes beyond a set threshold. The system has a text generation feature which generates texts either by keystrokes or by selecting words from a predefined list.

The first voluntary eye blink that is made helps the user to select their preferred text generation mode––the "Customized Message" panel with predefined text or the "Compose Text" panel by keystrokes.
In the customized message panel, message categories like \textit{home, office, hospital} and \textit{frequently used} texts are presented to the user and each category gains the cursor focus accordingly depending on the specified cursor speed. In this context, the cursor is a selector that hovers over each button/component in the interface.

A second blink transfers focus to a specific category and each message in the category gains focus after the third blink. The final eye blink selects the current text under focus, which is converted to an audio format and read out loud using the .NET speech synthesis engine. The text is copied automatically to the clipboard and the software switches to Skype window where the message is automatically pasted and sent to a recipient who has been pre-configured for a chat.

\subsection{Text generation}
A modification of the T9 algorithm (a predictive text algorithm) \cite{t9} is employed to support text suggestions. Instead of using 9 keys (as used in the original T9 algorithm), we use only 6 keys (T6) to generate texts; this significantly reduced the text generation time. This modified T9 algorithm employs trie data structure for its implementation. Trie is tree data structure that associates each node with a character and the words that terminate at that node. It allows \(O(n)\) traversal to find words matching an input (the depth of the tree corresponds with the length of the word).  
Texts suggestions in this implementation is the display of \(N\) number of words that have a T6 prefix matching the current input. For example, the T6 prefix for "apple" is 14431. Figure 1 is a diagram illustrating trie. Traversing from the root, the words THERE, THEIR, ANSWER, ANY, BYE can be constructed.

\subsubsection{Algorithm}
\begin{verbatim}
Step 1: Build a trie (add all words from dictionary to it).
Step 2:	Initially a current node is a root of the trie.
Step 3:	When a new character is generated, simply go to the next node
        from the current node by the edge that corresponds to this
        character (or report an error if there is nowhere to go).
Step 4:	Get all(or \(k\) first) possible words with a given prefix by 
        traversing the trie in breadth first search order 
        starting from the current node(if only k first words
        are needed, stop search when k words are found).
Step 5:	When the entire word is typed in and a new word 
        is started, move to the root again and repeat
        steps 3 - 5 for the next word.
\end{verbatim}
\begin{figure}
  \centering
  \includegraphics[width=0.27\linewidth]{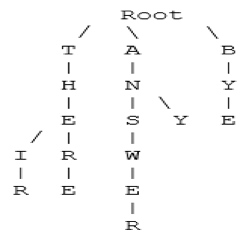}
  \caption{A trie for word generation }
  \label{fig:trie}
\end{figure}
All nodes that correspond to a word (not a prefix, but the entire word) is marked when the trie is built so that it is easy to understand whether a new word is found when traversing the trie at Step 4.\mbox{}\\
In the building process in Step 1, a list of all words represented by a certain number are mapped to that number. This is done for few possible combination of numbers and up to whatever length is desired, such that when the user types in a number, the program simply retrieves the list associated with that number or key. A file containing every possibility is compiled before the start of the program, this speeds up the search.

\subsubsection{Pseudo-code} Breadth-First Search (BFS) is a tree/graph traversal algorithm implemented via a queue. BFS visits all children nodes of a parent node before traversing in depth. This property is useful in text auto-complete because it allows same length words to be discovered before longer words. BFS algorithm is still currently widely applied in many domains \cite{bfs1, bfs2, bfs3, bfs4}.

In Step 4 of the T6 algorithm above, an adaptation of BFS is used to discover words.\mbox{}\\
\begin{verbatim}
T6BFSfunction(node):
   WHILE wordsList.length < numWords AND queue.length > 0 AND maxSearch > 0:
       	current = queue.dequeue()
        FOREACH current, FUNCTION(el, key): queue.push(el)
        IF current.words:
            wordsList = wordsList.concat(current.words)
        END IF
		        maxSearch = maxSearch - 1
    END WHILE
    RETURN wordsList
end
\end{verbatim}

\section{Results}
\label{others}
To establish connection with the headset, a user is required to select a COM port and baud rate (See Figure 2a). To configure the headset, the user chooses a cursor speed and a blink threshold value to determine how long each control gains focus and the strength of the blink required to determine a voluntary eye-blink, respectively (See Figure 2b). 

The application interface has 4 panels: the top panel (See top panel in Figure 2c) displays the generated texts or the current text being generated by the user; the left panel displays a list of pre-customized messages arranged by categories such as hospital, office, school and frequently used words (See left panel in Figure 2c). In practice, users will have the option to either select words from a list of frequently used words or generate their own words themselves. The user is then led to enter input through the simulated keypad shown on the interface (See center panel of figure 2c); a blink highlights a pad, and another blink selects the choice character when it is in focus. EEG readings are also displayed on the user interface (See right panel of Figure 2c).

\begin{figure}[h!]
  \centering
  \begin{subfigure}[b]{0.29\linewidth}
    \includegraphics[width=\linewidth]{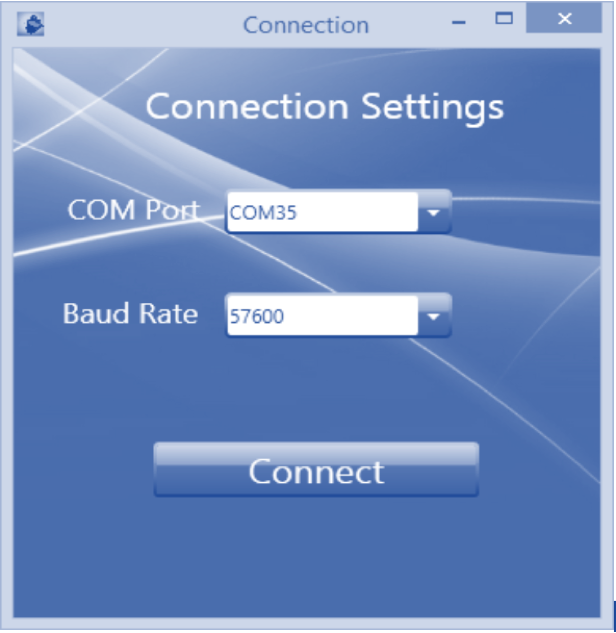}
    \caption{Connection interface}
  \end{subfigure}      
  \begin{subfigure}[b]{0.35\linewidth}
    \includegraphics[width=\linewidth]{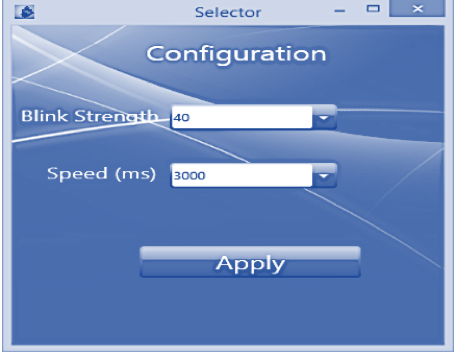}\hfill
    \caption{Blink threshold and selector speed interface}
  \end{subfigure}
  \begin{subfigure}[b]{1\linewidth}
    \includegraphics[width=\linewidth]{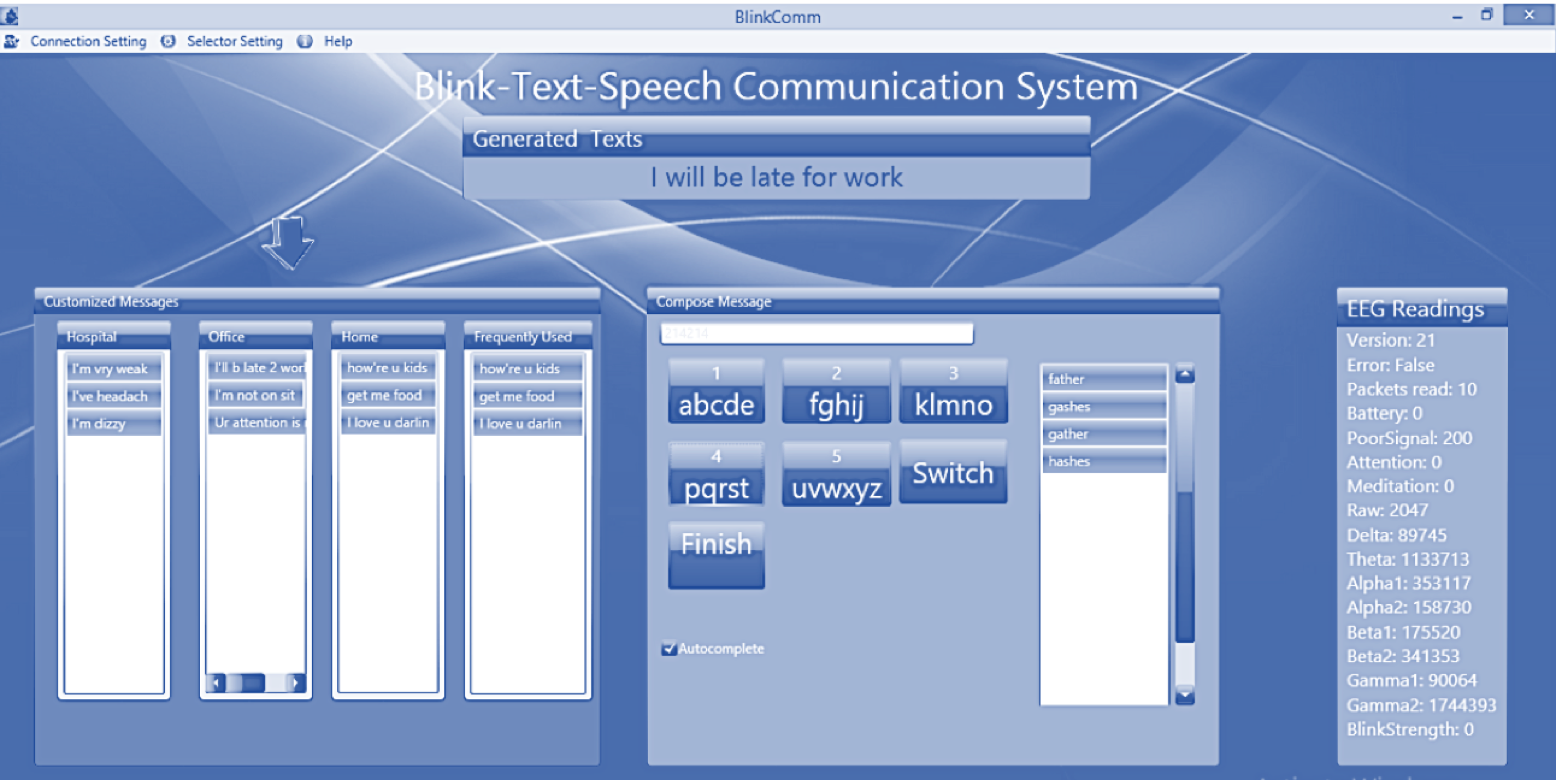}
    \caption{Text generation main interface }
  \end{subfigure}
  \caption{Application interfaces}
  \label{fig:setup}
\end{figure}

\section{Conclusion}
In this paper, we presented an electroencephalographic (EEG) based communication system which utilizes two EEG sensors to capture alpha values of EEG brain waves in real-time. These values are monitored and analysed to uncover voluntary eye-blink patterns. Upon detection of a pattern, a button is triggered on the software interface to generate texts from the user. Text generation is facilitated using the T9 algorithm––a predictive text algorithm used in mobile phones. Future work will address issues such as applying machine learning in the text prediction algorithm, designing the system configuration to learn user-centred and context-aware features, and adding messaging capabilities to the system.

\end{document}